\documentclass[aps,prl,twocolumn,groupedaddress,showpacs,showkeys]{revtex4-1}

\usepackage{graphicx}
\usepackage{float}
\usepackage{amsmath}
\usepackage{xfrac}
\usepackage{upgreek}
\usepackage{bm}
\usepackage[version=3]{mhchem}
\DeclareMathOperator\dif{d\!}
\newcommand\sbup{S_{\mathrm{\uparrow}}}
\newcommand\sbdn{S_{\mathrm{\downarrow}}}
\newcommand\sbco{S_{\mathrm{Co}}}
\newcommand\sbpt{S_{\mathrm{Pt}}}
\newcommand\sbn{S_n}
\newcommand\sgup{\sigma_{\mathrm{\uparrow}}}
\newcommand\sgdn{\sigma_{\mathrm{\downarrow}}}
\newcommand\sgco{\sigma_{\mathrm{Co}}}
\newcommand\sgpt{\sigma_{\mathrm{Pt}}}
\newcommand\pS{P_\mathrm{S}}
\newcommand\pT{P_{\mathrm{T}}}
\newcommand\dx{\nabla_x}

\newcommand\dt{\nabla T}
\newcommand\dxvs{\nabla_x V_{\mathrm{c}}}
\newcommand\dyvn{\nabla_y V_{\mathrm{N}}}
\newcommand\vs{V_\mathrm{c}}
\newcommand\vn{V_\mathrm{N}}
\newcommand\vh{V_\mathrm{H}}
\newcommand\js{J_\mathrm{s}}

\newcommand\jsx{J_{\mathrm{s},x}}

\newcommand\ha{\theta_{\mathrm{H}}}

\newcommand\spa{\theta_{\mathrm{s}}}
\newcommand\na{\theta_{\mathrm{N}}}
\newcommand\rxx{\rho_{xx}}
\newcommand\rxy{\rho_{xy}}
\newcommand\ns{\nu_\mathrm{c}}
\newcommand\nsn{\nu_{\mathrm{c},n}}
\newcommand\nsco{\nu_\mathrm{c,Co}}
\newcommand\nspt{\nu_\mathrm{c,Pt}}
\newcommand\nn{\nu_\mathrm{N}}
\newcommand\nnco{\nu_\mathrm{N,Co}}

\newcommand\unw{\ \mathrm{\upmu V/W}}
\newcommand\unk{\ \mathrm{\upmu V/K}}

\begin{document}


\title{Determining Spin Polarization of Seebeck Coefficients via Anomalous Nernst Effect}

\author{C. Fang}
\author{C.H. Wan}
\email{wancaihua@iphy.ac.cn}
\author{Z.H. Yuan}
\author{H. Wu}
\author{Q.T. Zhang}
\author{L.B. Mo}
\author{X. Zhang}
\author{X.F. Han}
\email{xfhan@iphy.ac.cn}

\affiliation{Institute of Physics, Chinese Academy of Sciences, Beijing, 100190, China}


\date{\today}

\begin{abstract}
Recently, Seebeck coefficients of ferromagnetic conductors are found to be spin-dependent. However straightforward method of accurately determining its spin polarization is still to be developed. Here, we have derived a linear dependence of anomalous Nernst coefficient on anomalous Hall angle with scaling factor related to spin polarization of Seebeck coefficient, which has been experimentally verified in [Co/Pt]$_n$ superlattices. Based on the dependence, we have also evaluated spin polarization of Seebeck coefficient of some ferromagnetic conductors. Besides, we have also found a new mechanism to generate pure spin current from temperature gradient in ferromagnetic/nonmagnetic hybrid system, which could improve efficiency from thermal energy to spin current.
\end{abstract}

\pacs{85.75.-d,72.15.Jf,72.20.Pa,85.80.-b}
\keywords{anomalous Nernst effect, Seebeck effect, pure spin current, spin orbit coupling, spin polarization of Seebeck coefficient, anomalous Hall effect}

\maketitle

Since the birth of spintronics, methods as nonlocal spin injection~\cite{ref01} , spin Hall effect~\cite{ref02,ref03}, spin pumping via ferromagnetic resonance~\cite{ref04,ref05} and circular polarized optical excitation~\cite{ref06} have been developed to generate pure spin current ($\js$), a long-aspired entity for its versatile functionality to realize magnetization rotation~\cite{ref07,ref08}, reversal~\cite{ref09,ref10,ref11} or domain wall motion~\cite{ref12}. Recently another novel category of methods, spin Seebeck effect (SSE)~\cite{ref13,ref14,ref15,ref16,ref17,ref18,ref19,wu2015} and spin-dependent Seebeck effect (SDSE)~\cite{ref14,ref20}, utilizing $\dt$ to produce pure $\js$ were demonstrated. $\dt$ leads to in a ferromagnetic (FM) layer out-of-equilibrium distribution of magnons which further damps back toward equilibrium state by emitting $\js$ into an adjacent nonmagnetic (NM) layer in SSE~\cite{ref14}. In SDSE~\cite{ref20}, instead, $\dt$ and resulting Seebeck voltage induced by the $\dt$ leads to, respectively, diffusion current and drift current. They not only have opposite directions but also different spin polarizations, thus generating a pure $\js$ depending on $\sbup-\sbdn$. Here $S_{\uparrow(\downarrow)}$ is Seebeck coefficient of spin up (spin down) carriers. In order to evaluate $\sbup-\sbdn$ as well as $\pS\equiv(\sbup-\sbdn)/(\sbup+\sbdn)$, Slachter~\cite{ref20} introduced the $\js$ into a NM material, measured spin accumulation in the NM material from which they further backward calculated $\js$. However, transparency of $\js$ from FM to NM layers is hard to evaluate accurately due to difficulty in determining spin mixing conductance~\cite{ref21} and some other interference effects such as interface spin memory loss effect~\cite{ref22} and magnetic proximity effect~\cite{ref19}. Here, instead, we have adopted another strategy, making use of inverse spin Hall effect (ISHE) of FM layer itself, to measure the $\js$ inside FM layer. We will also show in the following (1) that anomalous Nernst effect (ANE) whose scenario has long been in mystery has close relation with SDSE and anomalous Hall effect (AHE) and (2) that $\pS$ is obtainable by measuring ANE and Seebeck coefficient as well as anomalous Hall angle. These findings would be beneficial for developing a straightforward method to evaluate $\pS$ and thus also for searching materials with large transforming efficiency from heat flow to pure spin flow.

\begin{figure}[htb!]
\includegraphics[width=8.5cm]{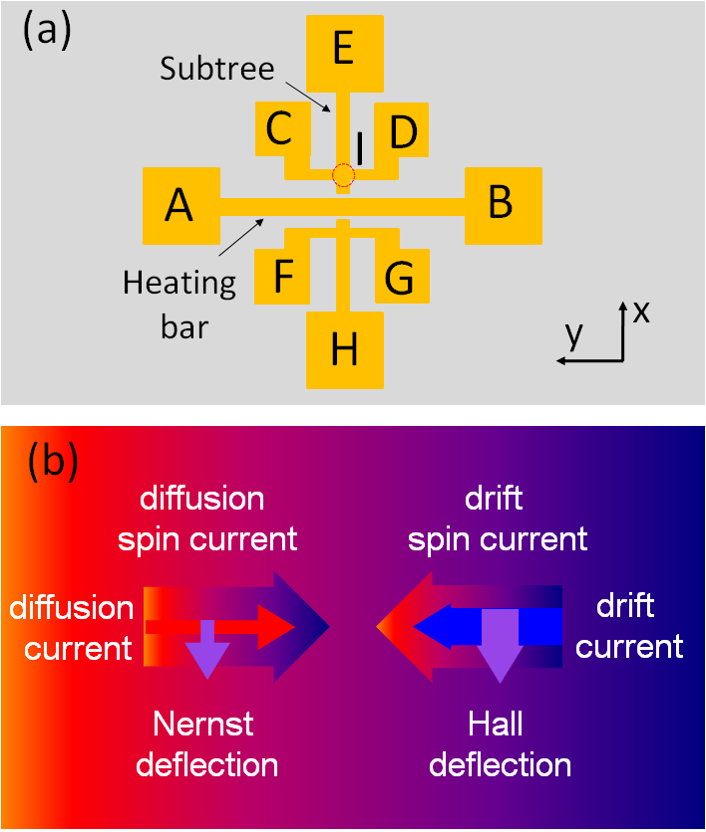}%
\caption{\label{fig1}Schematic diagrams of (a) measurement geometry and (b) physical scenarios of SDSE and ANE. Diffusion current is totally balanced by drift current while their spin counterparts is not, leading to a pure spin current along x and further a charge current along y due to ISHE in FM.}
\end{figure}

[Co($12/n$ nm)/Pt($12/n$ nm)]$_n$ with $n=1-8$ were magnetron-sputtered on \ce{Si}/\ce{SiO2} (500~nm) substrates with a base pressure of $1.0$ ($10^{-6}$ Pa) at room temperature. The number of Co/Pt interfaces increased as ($2n-1$) while their total thickness was kept as 12 nm. Co(12 nm), Pt(12 nm) and X(20 nm)/Pt(2 nm) (X=NiFe, CoFeB, Co, Fe) were fabricated for comparison. Then all films were patterned as Nernst Bar as shown in Fig.~\ref{fig1} (a). In Nernst measurement, heating current was transported between Pads A and B, resulting in a thermal gradient in transverse direction. In this direction, no current flowed between Pads C and E since the subtrees were isolated to the heating bar by $7\ \mathrm{\upmu m}$. Then Pads C and D were used to detect a pure Nernst voltage free of voltages due to any current related effect such as Peltier Effect or Anomalous Hall Effect. Pads C and E were used to detect Seebeck voltages. Magnetic field was provided by Physical Properties Measurement System (PPMS-9T, Quantum Design).

Spins in Co were first aligned along $z$ by a large $B_z$. $\dx T$ was generated by applying $I$ between Pads A and B. Seebeck voltage $\vs$ was then built. $\dx \vs=(\sgup\sbup+\sgdn\sbdn)/\sgco\dt=\sbco\dx T$. Here $\sgup$ , $\sgdn$ and $\sgco=\sgup+\sgdn$ were conductivity of spin up, spin down channels and total conductivity, respectively. Although diffusion current $\sgco\sbco\dx T$ was fully balanced by drift current $\sgco\dx\vs$, diffusion spin current $\pT\sgco\sbco\dx T$ differed from drift spin current $P\sgco\dxvs$ due to their different spin polarizations, thus $\js$ emerging with $\jsx=(P-\pT)\sgco\dxvs$ (SI A) (Fig.~\ref{fig1} (b)). Spin polarization $P$ of conductivity, spin polarization $\pT$ of thermoelectric conductivity ($\sigma S$) and spin polarization $\pS$ of Seebeck coefficient were defined as $P\equiv(\sgup-\sgdn)/\sgco$, $\pT\equiv(\sgup\sbup-\sgdn\sbdn)/(\sgup\sbup+\sgdn\sbdn)$ and $\pS\equiv(\sbup-\sbdn)/(\sbup+\sbdn)$, respectively. They correlated via $\pS=(\pT-P)/(1-\pT P)$. The derived $\js$ was the same as derived by Slachter~\cite{ref20}. They further introduced $\js$ into a NM material and measured it via spin accumulation in the NM material while we calibrated it by ISHE in FM itself directly (SI A). Resulted voltage $\vn$ (ANE voltage) was thus
\begin{equation}
\dyvn=(1-\pT/P)\ha\dxvs
\end{equation}

ANE coefficient $\eta\equiv\dyvn/(\mu_0M_{0z}\dx T)$, $\mu_0$ and $M_0$ permeability of vacuum and saturated magnetization, respectively. Anomalous Hall angle $\ha$, ANE coefficient $\eta$ and Seebeck coefficient $\sbco$ were correlated via Eq.\ (2).

\begin{equation}
\eta=(1-\pT/P)\ha\sbco/(\mu_0M_{0z})
\end{equation}

Here $\ha=P\spa$ where $\spa$, spin Hall angle, characterized spin-orbit coupling (SOC) strength of FM material. Linear scalability of $\eta$ with $\ha$ and $S$ was obtained previously in nonmagnetic materials~\cite{ref23} but not in ferromagnetic counterparts. It deserved special attention that quantity $(\dyvn/\dx T-\ha\sbco)$ which was lack of deep comprehension in previous literatures~\cite{ref24,ref25} became interpretable after $\pT$ was introduced. Similar with $\ha$, Nernst angle $\na=\pT\spa$.

Multilayer differed subtly from a single layer in thermotransport properties. Taking Pt and Co bilayer as an example and supposing they had the same thickness, $S_\mathrm{eff}=(\sgco\sbco+\sgpt\sbpt)/(\sgco+\sgpt)$. $\jsx$ in Co of the bilayer was remarkably modified as $\jsx=\sgco/(\sgco+\sgpt)\sbco(P-\pT)\sgco\dx T+(P\sbpt-\pT\sbco)\sgco\sgpt/(\sgco+\sgpt)\dx T$ (SI B). There existed two origins contributing to $\jsx$, the $1^\mathrm{st}$ term from intrinsic SDSE but reduced by a factor $\sgco/(\sgco+\sgpt)$ due to shielding effect of Pt and the $2^\mathrm{nd}$ term named as extrinsic SDSE from variation of Seebeck coefficients in adjacent layers. To highlight the extrinsic one, one could suppose a fictional system where intrinsic SDSE vanished ($\pT=P$). Then $\jsx=-P(\sbco-\sbpt)\sgco\sgpt/(\sgco+\sgpt)\dx T$. In contrast, $\jsx=-2(\sbup-\sbdn)\sgup\sgdn/(\sgup+\sgdn)\dx T$ for a single FM layer. If Co and Pt in the bilayer were deemed as two spin channels with $P$ and 0 spin polarizations, respectively, the bilayer behaved similarly to a FM ¡°single¡± layer. SOC considered, spin current also introduced a transverse ANE voltage via Eq.\ (3).
\begin{widetext}
\begin{equation}
\frac{\dyvn}{(\ha\dxvs)}=\frac{(1-\pT/P)\sgco/(\sgco+\sgpt)+(\sbpt/\sbco-\pT/P)\sgpt/(\sgco+\sgpt)}{1+(\sgpt\sbpt)/(\sgco\sbco)}
\end{equation}
\end{widetext}

The above analysis showed that ANE, AHE and SDSE were correlated with each other: SDSE in ferromagnetic materials generated pure spin current which led to a transverse ANE voltage through SOC in the same manner as occurred in AHE. It also indicated that $\pT/P$ was obtainable from anomalous Nernst effect.

We have prepared pure Co films (12 nm) and [Co($12/n$ nm)/Pt($12/n$ nm)]$_n$ superlattices to modulate SOC strength, then measured their AHE, Seebeck effect (SE) and ANE in order to measure $\pT/P$ of pure Co as well as to testify existence of the extrinsic SDSE.\@ Practically, Seebeck voltage and Nernst voltage were measured between Pads C--E and Pads C--D, respectively (Fig.~\ref{fig1} (a)). To remove a same parameter $(\nabla_\mathrm{I}T)/[(T_\mathrm{I}-T_\mathrm{E})/r_\mathrm{IE}]$ introduced by nonlinear $\dx T$ (SI C), Eq.\ (3) would be divided by Eq.\ (1). Here $(\nabla_\mathrm{I}T)$ is temperature gradient at Point I while $(T_\mathrm{I}-T_\mathrm{E})/r_\mathrm{IE}$ is average temperature gradient between Point I and Pad E.\@

\begin{figure}[htb!]
\includegraphics[width=8.5cm]{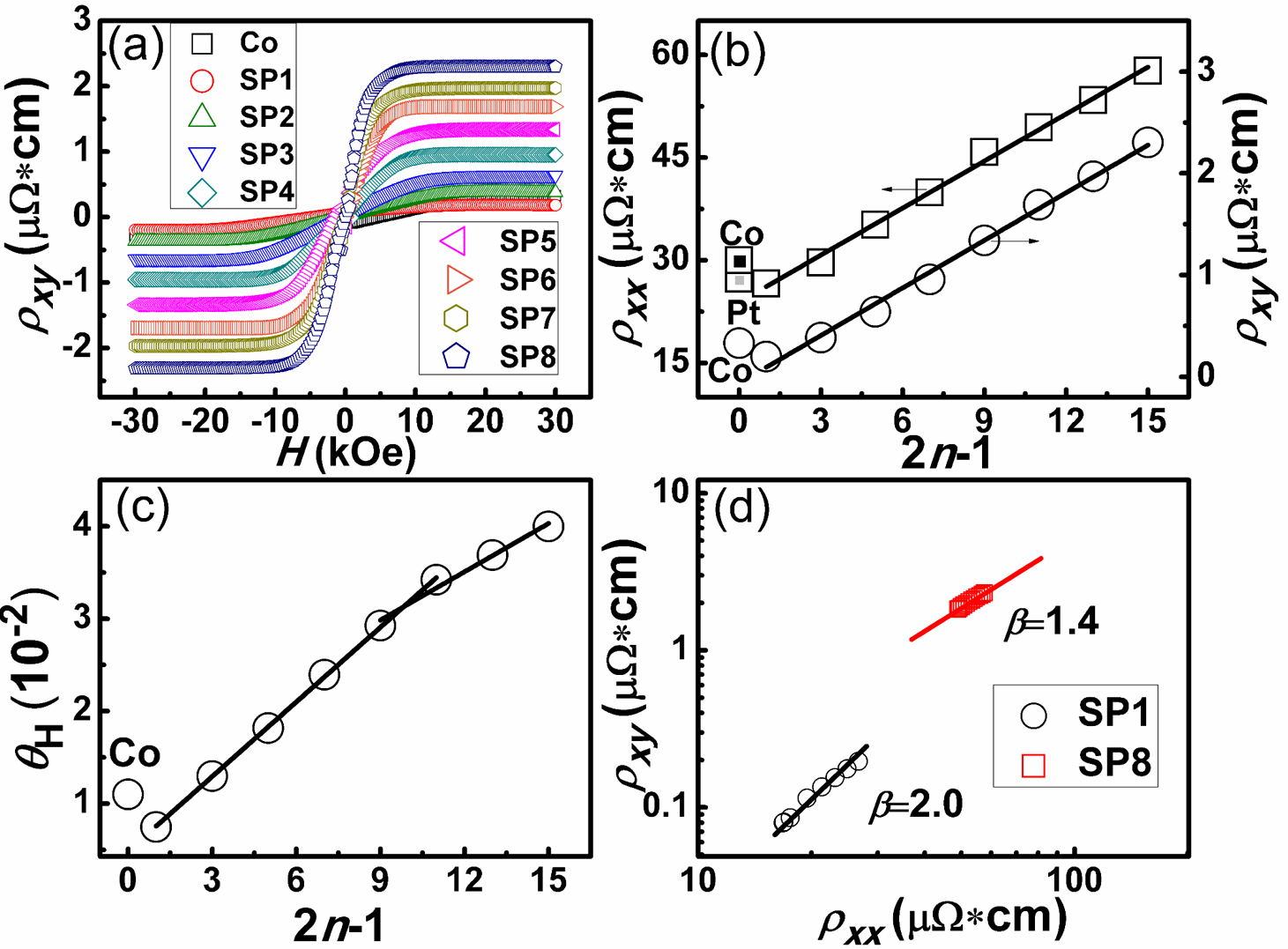}%
\caption{\label{fig2}(a) $B$ dependence of $\rxy$, (b) $\rxx$ and $\rxy$ of SP$n$, (c) $\ha$ of SP$n$ at 300 K and (d) scaling law between $\rxy$ and $\rxx$ of SP$1$ and SP$8$ from 300 K to 10 K. The black and red lines show linear fitting of $\rxy$ vs.\  $\rxx$, respectively.}
\end{figure}

Fig.~\ref{fig2} (a) showed $B$ dependence of $\rxy$ of the samples (SP$n$, $n=1-8$) at 300 K.\@ 2.5 T Field was high enough to saturate $M$ along $z$. $\rxy$ and $\rxx$ both linearly depended on the number of Co/Pt interfaces ($2n-1$) (Fig.~\ref{fig2} (b)), indicating interfacial scattering and extrinsic skew scattering gradually dominated $\rxx$ and $\rxy$, respectively. $\ha$ of SP$n$ was summarized in Fig.~\ref{fig2} (c). $\ha$ of SP$8$ was 0.040, much larger than 0.0075 of SP$1$ and 0.011 of pure Co. Scaling law $\rxy\propto{\rho}_{xx}^{\phantom{xx}\beta}$ was measured by varing $T$. Scaling exponent $\beta$ of SP$1$ was 2 while that of SP$8$ was reduced to 1.4 (Fig.~\ref{fig2} (d)), also demonstrating dominating role of skew scattering in samples with larger $n$~\cite{ref26}. Similar results that Co/Pt interfaces induced skew scattering were also reported by Canedy~\cite{ref27}.

Seebeck coefficients of pure Co (12~nm), Pt (12~nm) and the superlattices were also measured with heating $I$ flowing through Pads A and B and voltmeter connecting Pads C and E (SI D). $\dx T$ was proportional to heating power $P$ . Thus $\ns\equiv\dif \vs/\dif P$ and $\nn\equiv\dif \vn/\dif P$ were used to indirectly characterize $S$ and $\eta$. $\nspt=-(7.8\pm0.4)\unw$ and $\nsco=-(23\pm2)\unw$. $\sbpt=-5.0\unk$~\cite{ref28}. Then $\sbco=-(14.7\pm1.5)\unk$, close to $-22\unk$ applied in Ref.~29\nocite{ref29}. The $\nsn$ exhibited no systematic dependence on $n$. Averagely $\nsn=(16\pm3.0)\unw$. Accordingly, $\sbn=-(10.3\pm2.0)\unk$. The formula $\sbn=(\sgco\sbco+\sgpt\sbpt)/(\sgco+\sgpt)$ predicted $\sbn\cong-(10.1\pm0.8)\unk$ considering $\sgpt/\sgco\cong1.1$, consistent with the measured value. $T_\mathrm{I}-T_\mathrm{E}$ and average $\dx T$ between Point I and Pad E were thus 0.78 K and 22 K/cm at $P=0.5~W$. Even higher $\dx T$ existed in regions closer to the heating bar (SI C) where giant ANE was expected. Due to effectiveness to produce large $\dt$, this geometry has also been applied to study thermoelectric properties of carbon nanotubes~\cite{ref30} and graphene~\cite{ref31}.

\begin{figure}[htb!]
\includegraphics[width=8.5cm]{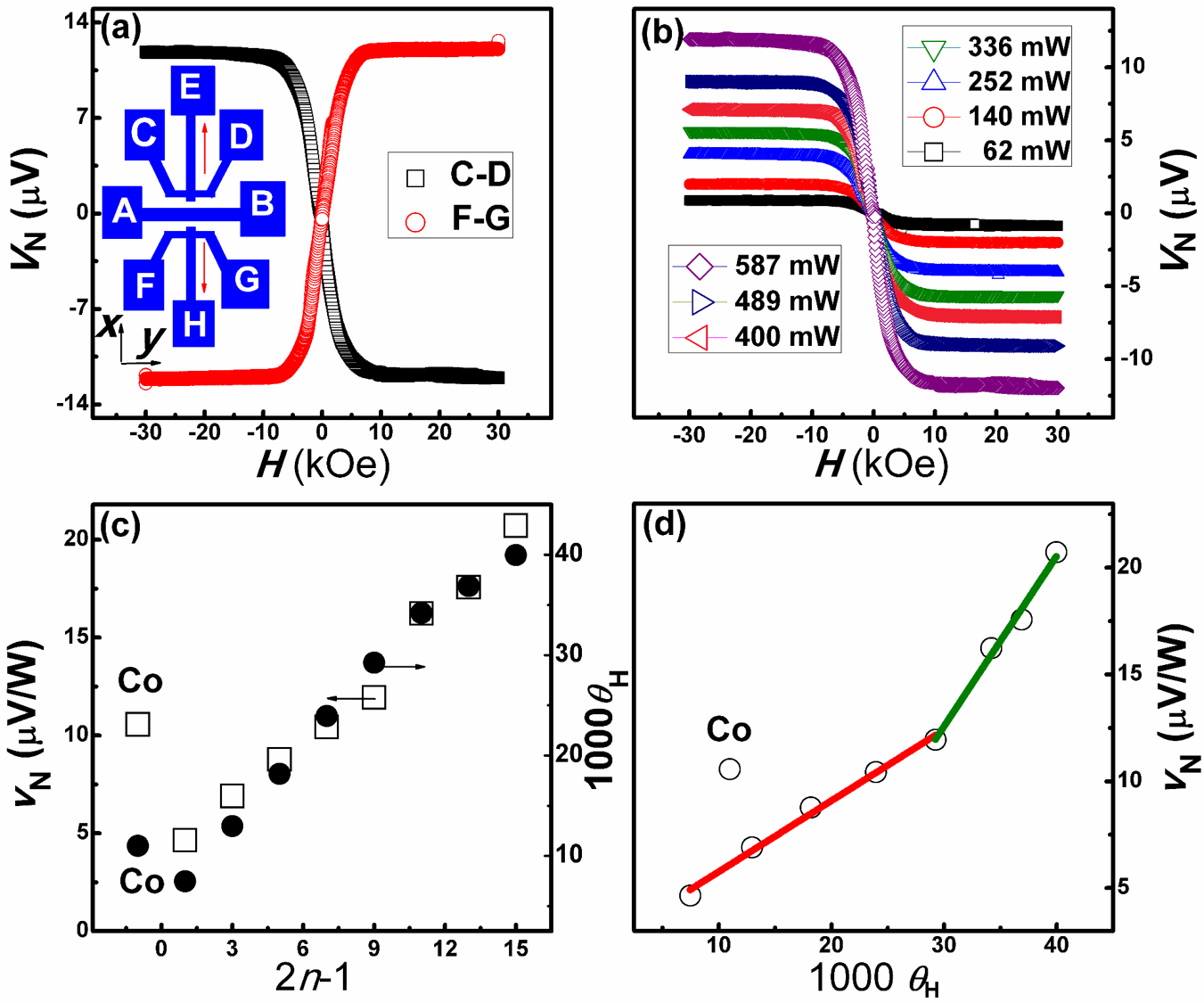}%
\caption{\label{fig3}(a) $\vn$ between Pads CD and FG, inset showing corresponding setups, (b) the dependences of $\vn$ between CD on $B$ under different $P$, (c) the dependence of $\nn$ and $\ha$ on $2n-1$ and (d) the scaling relation between $nn$ and $\ha$, the lines representing linear fittings of $\nn$ vs.\ $\ha$ as $n\le5$ and $n>5$.}
\end{figure}

$\vn$ evolved with $B_z$ in a similar manner with $\vh$ and sign of $\vn$ was reversed as $\dx T$ was reversed, a typical feature of ANE (Fig.~\ref{fig3} (a)). Fig.~\ref{fig3} (a-b) also showed ordinary Nernst effect was small enough to be ignored up to 3 T\@. Saturated $\vn$ increased linearly with $P$ as expected (Fig.~\ref{fig3} (b)), thus $\nn\equiv\dif\vn/\dif P$ did not depended on $P$. SP$8$ exhibited much higher $\nn$ than SP$1$ at the same condition. Similar with linear dependence of $\ha$ on $n$ as $n<6$, $\nn$ of the superlattices also linearly increased with $n$ (Fig.~\ref{fig3} (c)). $\nu_\mathrm{N,5}=2.6\nu_\mathrm{N,1}$, indicating interfacial SOC was the common origin of AHE and ANE (Fig.~\ref{fig3} (d)). This linear relation also validated Eq.\ (1) and Eq.\ (3). Though dependencies of $\ha$ and $\nn$ on $n$ deviated linear relation beyond $n\ge5$, $\nn$ still linearly scaled with $\ha$ as $n\ge5$ with a larger $\dif\nn/\dif\ha$. $\dif{\nn}/\dif\ha =(0.33\pm0.02)~\mathrm{mV/W}$ and $(0.80\pm0.06)~\mathrm{mV/W}$, respectively, as $n\le5$ and $n\ge5$.

$C_\mathrm{0}\equiv\nnco/(\ha\nsco)=42\pm4$ for the Co single layer while $C_n\equiv(\dif{\nn}/\dif{\ha})/\nsn=21\pm4$ as $n\le5$, averagely. As indicated by Eq.\ (3), $C_n$ should have been about $40\% C_\mathrm{0}=17\pm2$ as $\sgpt\cong0.91\sgco$ and $\sbpt/\sbco=0.34$ if only the intrinsic SDSE contributed to spin current. The measured $C_n(n\le5)$ was $(1.24\pm0.28)$ times as large as the estimated value, indicating the intrinsic SDSE played major role. In contrast, $C_n=50\pm10$ as $n\ge5$, two times larger than the expected value. The extra part, as per our knowledge, has to be attributed to the extrinsic SDSE\@. By calculating [Eq.\ (3)/ Eq.\ (1)] with $C_n=50\pm10$ and $C_\mathrm{0}=42\pm4$, one could remove the parameter $(\nabla_\mathrm{I}T)/[(T_\mathrm{I}-T_\mathrm{E})/r_\mathrm{IE}]$ (SI C) and then obtain $\pT/P=1.56\pm0.35$ and $\pT=-(0.62\pm0.14)$ if $P=-0.4$~\cite{ref32}. According to the results in Ref.\ 29\nocite{ref29}, $\pT$ was estimated as $51\%$, also close to the result measured here. According to Eq.\ (2), $\eta\cong-50\ \mathrm{nV/(K\ T)}$ considering $\mu_0M_0=1.7\ \mathrm{T}$,  in the same order of $\eta$ of \ce{Co2FeAl}~\cite{ref33}. $\pS=-(0.30\pm0.20)$ for pure Co and $\dx T$ at Nernst cross was about $0.165~\mathrm{K/\upmu m}$ at $P=0.5\ \mathrm{W}$ and $(\nabla_\mathrm{I}T)/[(T_\mathrm{I}-T_\mathrm{E})/r_\mathrm{IE}]$ was about 75. For a pure Co layer, $\dif\jsx/\dif\dx T\cong(11\pm7.5)\ \mathrm{A/(m\ K)}$ therefore, $\jsx=(1.8\pm1.2)\ \mathrm{MA/m^2}$ as $P=0.5\ \mathrm{W}$, far enough to drive rotational and translational motions of skyrmions~\cite{ref34}. It was noteworthy that the extrinsic SDSE arose as $n\ge5$ or spacing between Pt layers became shorter than 2.4 nm, hinting the extrinsic SDSE was merely remarkable in interfacial regions by less than 1.2 nm. The length scale is comparable to typical spin diffusion length in Pt. Compared with the intrinsic SDSE, the extrinsic SDSE could be more flexibly modulated by suitable choice of materials whose Seebeck coefficients differed by large values.

\begin{figure}[htb!]
\includegraphics[width=8.5cm]{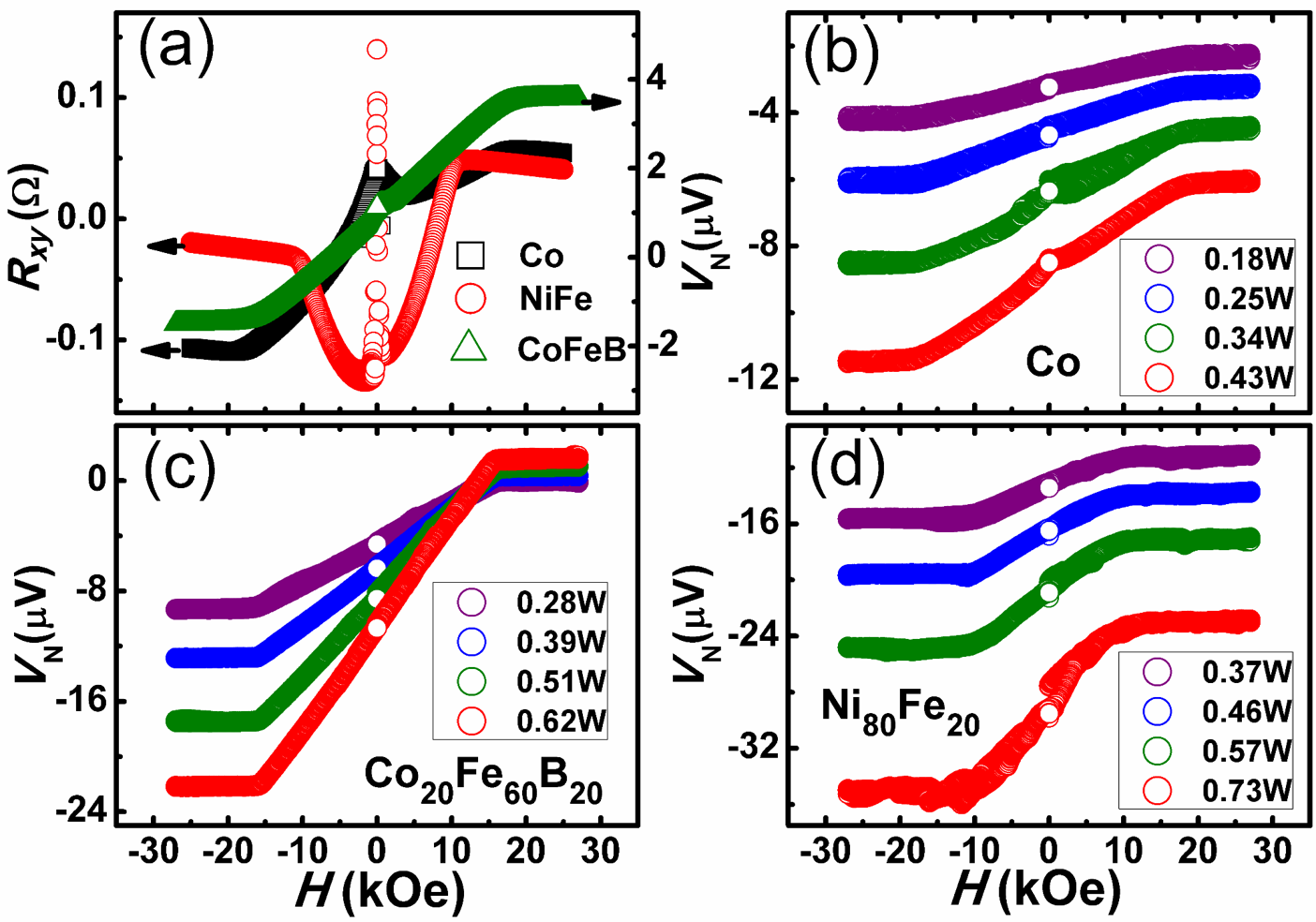}%
\caption{\label{fig4}(a) AHE and (b, c and d) ANE between Pads F--G of X (Co, \ce{Co40Fe40B20}, and \ce{Ni80Fe20}, 20 nm)/Pt(2 nm) at elevated P.}
\end{figure}

We also investigated $\pT/P$ of some other ferromagnetic materials as \ce{Co40Fe40B20}, \ce{Ni80Fe20} (Fig.~\ref{fig4}) and Fe (not shown here). $\eta_{Fe}$ was negligibly small due to very small and positive $S_\mathrm{Fe}$. Structures of X (20 nm)/Pt (2 nm) were deposited where X=Co, \ce{Co40Fe40B20} and \ce{Ni80Fe20}. Pt layer functioned as anti-oxidation layer. The $\ns$ of Co, \ce{Co40Fe40B20} or \ce{Ni80Fe20} were $-19.4\unw$, $-7.78\unw$ and $-16.1\unw$, respectively. Their $\nn$ were $7.27\unw$, $21.7\unw$ and $11.3\unw$, respectively. Their Hall angles were 0.011, 0.048 and 0.0032, respectively. Thus their $|C_\mathrm{X/Pt}|$ were about 34, 58 and 220, respectively. $(\nabla_\mathrm{I}T)/[(T_\mathrm{I}-T_\mathrm{E})/r_\mathrm{IE}]=75$ taken into account, their $\pT/P$ were accordingly about 1.45, 1.77 and 3.93. If $|P|$ of Co, \ce{Co40Fe40B20} and \ce{Ni80Fe20} were taken as 0.4~\cite{ref32}, 0.5~\cite{ref35} and 0.2~\cite{ref36}, their $|\pT|$ were about 0.58, 0.89 and 0.79, respectively and their $\pS$ were about 0.23, 0.70 and 0.70.  $|P_\mathrm{T,Co/Pt}|=0.58$ was comparable to $|P_\mathrm{T,Co}|=0.62\pm0.14$ measured in the superlattices. \ce{Ni80Fe20} and \ce{Co40Fe40B20} had higher $\pT$ and $\pS$ than Co. The result that \ce{Ni80Fe20} had larger $\pS$ than Co has also been reported in Ref.~29\nocite{ref29}, though different in magnitudes.

In summary, ANE was interpreted as joint action of SDSE and SOC. ANE, AHE and SDSE were then combined to characterize $\pT$ and $\pS$ in ferromagnetic films. For cobalt, $\pT=-(0.62\pm0.14)$ and $\pS=-(0.30\pm0.20)$ and $\dif \jsx/\dif \dx T\cong(11\pm7.5)\ \mathrm{A/(m\ K)}$. For \ce{Co40Fe40B20} and \ce{Ni80Fe20}, $\pT/P$ was about 1.77 and 3.93, respectively. ANE could provide a straightforward method to calibrate spin polarizations of Seebeck coefficients of ferromagnetic conductors. We also distilled an extrinsic SDSE originating from variation of Seebeck coefficients. It only significantly affected spin transport within interfacial regions of less than 1.2 nm for Co/Pt interface. This extrinsic mechanism could offer another flexible strategy to generate pure spin current via $\dt$, comparable to intrinsic mechanism, which could be applied especially in skyrmion spin-caloritronics due to its moderate efficiency from heat to spin current.

\begin{acknowledgments}
This research has been supported by the MOST National Key Scientific Instrument and Equipment Development Projects [No./ 2011YQ120053], National Natural Science Foundation of China [NSFC, Grant No./ 11434014 and 51229101], Natural Science Foundation for the Youth (Grant No./ 11404382) and Postdoctoral Science Foundation of China (Grant No./ 2013M540154).
\end{acknowledgments}

%

\end{document}


\title{Supplementary Information}
\maketitle
\setcounter{equation}{0}
\renewcommand{\theequation}{S\arabic{equation}}
\setcounter{figure}{0}
\renewcommand{\thefigure}{S\arabic{figure}}

Supplementary Information (SI) for Manuscript ¡°Determining Spin Polarization of Seebeck Coefficients via Anomalous Nernst Effect¡± by C. Fang, C.H. Wan *, Z.H. Yuan, H. Wu, Q.T. Zhang, L.B. Mo, X. Zhang, X.F. Han$ ^\dag$ from Institute of Physics, Chinese Academy of Sciences, Beijing, 100190, China.
\\ \\
\textbf{Outlines of Supplementary Information}\\
A.	Derivation of relation between ANE coefficient, AHE angle and Seebeck coefficient in single ferromagnetic layer system\\
B.	Derivation of relation between ANE coefficient, AHE angle and Seebeck coefficient in ferromagnetic/nonmagnetic bilayer system\\
C.	Estimation of temperature distribution\\
D.  Dependence of Seebeck voltages on heating power\\
E.	Magnetic and magnetotransport properties of [Co/Pt]$_n$ multilayers and pure Co layer\\ \\
\textbf{A. Derivation of relation between ANE coefficient, AHE angle and Seebeck coefficient in single ferromagnetic layer system}

  We will follow the semiotic system in Ref.\ \cite{Bauer2012} in the following discussion. According to diffusive and linear response theory~\cite{Bauer2012}, charge current density $\jc$, spin current $\js$ and heat flow $Q$ are driven by electric potential gradient $\nabla \mu _c$, spin potential gradient $\nabla \mu _s$ and temperature gradient $\dt$, according to Eq.\ (S1).
  \begin{equation}
  \begin{bmatrix}
  \jc \\
  \js \\
  Q
  \end{bmatrix}
  =\sigma
  \begin{bmatrix}
  1 & P & ST\\
  P & 1 & \pT ST\\
  ST& \pT ST& \kappa T/\sigma
  \end{bmatrix}
  \begin{bmatrix}
  \nabla \mu _c/e \\
  \nabla \mu _s/2e \\
  -\nabla T/T
  \end{bmatrix}
  \end{equation}
  Here $J_{c/s}=J_\uparrow\pm J_\downarrow$ where $J_\uparrow$ and $J_\downarrow$ are current density in spin up ($\uparrow$, defined according to direction of magnetization) and spin down ($\downarrow$) channels. $S$, $\sigma$ and $\kappa$ are total Seebeck coefficient, total conductivity and total thermal conductivity of a ferromagnetic material, respectively. $S=(\sbup\sgup+\sbdn\sgdn)/(\sgup+\sgdn)$ and $\sigma=\sgup+\sgdn$. $P=(\sgup-\sgdn)/(\sgup+\sgdn)$. $\pT=[{\partial(P\sigma)}/{\partial\epsilon}|_{\epsilon_\mathrm{F}}]/[{\partial\sigma}/{\partial\epsilon}|_{\epsilon_\mathrm{F}}]$. $\sgup$ and $\sgdn$ are spin-dependent conductivities. Electrochemical potential $\mc=(\mup+\mdn)/2$ and spin accumulation $\ms=(\mup-\mdn)/2$ where $\mup$ and $\mdn$ are electrochemical potential for spin up and spin down electrons. Compared with $\mc$, $\ms$ is relatively small in bulk region as implemented too as a hypothesis in Ref.\ \cite{ramos2014,slachter2010}. Then we arrive at Eq.\ (S2). $Q$ is ignored in Eq.\ (S2) since heat flow driven by $\nabla\mc$ and $\nabla\ms$ in our experiment is much smaller than what we have applied by an external heating current of about several tens of milliamps. In other words, $\dt$ is dominantly determined by external heating current instead of by $\nabla\mc$ and $\nabla\ms$.
  \begin{equation}
  \begin{bmatrix}
  \jc \\
  \js
  \end{bmatrix}
  =\sigma
  \begin{bmatrix}
  \nabla \mc/e - S\dt \\
  P\nabla \mc/e - \pT S\dt
  \end{bmatrix}
  \end{equation}
  For an open-circuited film, $\jc=0$, thus $\nabla\mc=eS\dt$, $\mc/e$ being Seebeck voltage actually in this case and Eq.\ (S3) holds.
  \begin{equation}
  \js=(P-\pT)\sigma S\dt=(P-\pT)\sigma\nabla\mc/e
  \end{equation}
  Interestingly, spin current $\js$ still survives even though net charge current $\jc$ in this case does not exist if $P\neq\pT$, which is the reason why the $\js$ is named as a pure spin current. Actually, $\pT=(\sgup\sbup-\sgdn\sbdn)/(\sgup\sbup+\sgdn\sbdn)$ taking Mott relation $\sigma_iS_i=\frac{\pi^2}{3}(\frac{k}{e})(kT)\frac{\dif\sigma_i}{\dif\epsilon}$ into account where $k$ is Boltzmann constant. $\pT$ is spin polarization of thermoelectric conductivity ($\alpha=\sigma S$). Thus Eq.\ (S3) could also be equivalent to equation (S4), which is also obtained in Ref.\ \cite{slachter2010}.
  \begin{equation}
  \js=-(1-P^2)/2(\sbup-\sbdn)\sigma\dt
  \end{equation}
  The above equation tells us (1) that pure spin current originates from difference between $\sbup$ and $\sbdn$ and (2) that spin polarization of Seebeck coefficient $\pS\equiv(\sbup-\sbdn)/(\sbup+\sbdn)$ could be obtained if the pure spin current $\js$ generated by $\dt$ could be measured. The relation among $\pT$, $\pS$ and $P$ is as shown in Eq.\ (S5).
  \begin{equation}
  \pT=\frac{\pS+P}{1+\pS P}
  \end{equation}
  Pure spin current vanishes only if $\sbup=\sbdn$ or $\pS=0$ or $\pT=P$.

  In this paper, in order to detect $\js$, we have adopted spin-orbit coupling (SOC) in ferromagnetic material (FM) itself to transform the $\js$ into a charge current $\jc$. Before going deep into anomalous Nernst effect, we first discuss how a longitudinal $\js$ induces a transverse $\jc$ by anomalous Hall effect (AHE). Supposing current $J$ is applied in FM along $x$, the $J$ could be decomposed into two spin components: $J(1+P)/2$ for spin-up electrons and $J(1-P)/2$ for spin-down electrons. Thus a spin current $PJ$ is produced. Without loss of any generality, we further suppose electrons with different spins could have (but not necessarily) different spin Hall angles $\theta_{\uparrow/\downarrow}$. $P_\theta\equiv(\thup-\thdn)/(\thup+\thdn)$ and let average spin Hall angle $\spa\equiv(\thup+\thdn)/2$. Thus a transverse current $\jc=\thup J(1+P)/2-\thdn J(1-P)/2=\spa J(P+P_\theta)$ is generated. Anomalous Hall angle $\ha=\jc/J=\spa(P+P_\theta)$.

  As occurred in AHE, spin current generated by $\dt$ in spin-dependent Seebeck effect (SDSE) could also be transformed as a transverse charge current through SOC, leading to anomalous Nernst effect (ANE). However, different from AHE, $\js$ in SDSE is a pure spin current which leads to a transverse current of $\jc=\thup\js/2+\thdn\js/2=\spa\js$. Ratio of $\jc/\js=\spa$ in this case, deviating from $\ha$ by a factor of ($P+P_\theta$).

  Considering (S3), the induced Nernst voltage ($\vn$), in open circuit condition, is then
  \begin{equation}
  \nabla\vn=\spa\js/\sigma=(P-\pT)\spa\nabla \mc/e
  \end{equation}
  It is worth noting that anomalous Hall angle $\ha$ equals to $\spa(P+P_\theta)$. Thus
  \begin{align*}
  \nabla\vn & =\frac{P-\pT}{P+P_\theta}\ha\nabla \mc/e \\
  C & \equiv \frac{\nabla\vn}{\ha\nabla (\mc/e)}=\frac{P-\pT}{P+P_\theta}
  \end{align*}
  Anomalous Nernst voltage $\vn$ proportions product of $\ha$ and $\nabla\mc/e$ with scaling factor $C$ depending on spin polarization of conductivity, thermoelectric conductivity and spin Hall angles in FM. Here $\thup=\thdn$ or $P_\theta=0$ is supposed because (1) we think electrons with different spins experience the same scattering potential in FM and (2) there have been few experimental or theoretical literatures to show their difference according to our knowledge. Thus
  \begin{subequations}
  \begin{align}
  \nabla\vn & =\frac{P-\pT}{P}\ha\nabla \mc/e \\
  C & \equiv \frac{\nabla\vn}{\ha\nabla(\mc/e)} =\frac{P-\pT}{P}
  \end{align}
  \end{subequations}
The above results could also be derived from another formalism as adopted in Ref.\ \cite{ramos2014,pu2008}. Supposing $J_i=\sigma_{ij}E_j-\alpha_{ik}\nabla_kT$, $\sigma$ and $\alpha$ being conductivity and thermoelectric conductivity tensors, respectively. $J_i$ and $E_j$ are charge current density along $i$ and electric field along $j$, respectively. Under insulative boundary conditions and $\dy T=0$,
\begin{equation}
\begin{bmatrix}
E_x \\
E_y
\end{bmatrix}
=\nabla_x T\begin{bmatrix}
\rxx\alpha_{xx}+\rxy\alpha_{xy} \\
\rxy\alpha_{xx}-\rxx\alpha_{xy}
\end{bmatrix}
\end{equation}
$\rho$ is resistivity tensor.\\
Seebeck tensor is defined as usual as $S_{ij}\equiv E_i/\nabla_jT$ . Then
\begin{equation}
\begin{bmatrix}
S_{xx} \\
S_{yx}
\end{bmatrix}
=\begin{bmatrix}
\rxx\alpha_{xx}+\rxy\alpha_{xy} \\
\rxy\alpha_{xx}-\rxx\alpha_{xy}
\end{bmatrix}
\end{equation}
Hall angle $\tan\ha\equiv\rxy/\rxx$ . For convenience, we define another angle, Nernst angle, as $\tan\na\equiv\alpha_{xy}/\alpha_{xx}$ . Then
\begin{equation*}
\begin{bmatrix}
S_{xx} \\
S_{yx}
\end{bmatrix}
=\rxx\alpha_{xx}\begin{bmatrix}
1+\tan\ha\tan\na \\
\tan\ha-\tan\na
\end{bmatrix}
\end{equation*}
Bear in mind that $\ha$ and $\na$ are both much smaller than 1 for majority ferromagnetic materials, thus $S_{xx}\cong \rxx\alpha_{xx}$ and $S_{yx}\cong S_{xx}(\tan\ha-\tan\na)$. The Nernst angle $\na$ reflects SOC induced deflection of \textit{diffusion current} under $\nabla T$. Similar with $\ha=P\spa$, $\na=\pT\spa$. Here, $\pT$ is spin polarization of thermoelectric conductivity defined above. Thus $\na/\ha=\pT/P$. Then $S_{yx}\cong S_{xx}(\ha-\na)=S_{xx}\ha(1-\pT/P)$ as shown in Eq.\ (S7). Here we also suppose $P_\theta=0$.\\ \\
\textbf{B. Derivation of relation between ANE coefficient, AHE angle and Seebeck coefficient in ferromagnetic/nonmagnetic bilayer system}

Thermal transport properties of ferromagnetic and nonmagnetic bilayer system show some subtle differences from the case of the single ferromagnetic layer. The bilayers are restricted to have the same thickness $h$ and the same width $w$ for the ferromagnetic and nonmagnetic layer for simplicity. In this system, charge and spin currents satisfy Eq.\ (S10). Here subscript Pt and Co are used to label quantities in Pt and Co layer respectively. Subscripts $\uparrow$ and $\downarrow$ also label quantities in Co layer. $\vs\equiv\nabla\mu/e$ Seebeck voltage.
\begin{widetext}
\begin{align}
\jc&=-\sgup\sbup\nabla T-\sgdn\sbdn\nabla T-\sgpt\sbpt\nabla T+(\sgco+\sgpt)\nabla\vs=0 \notag \\
\jsco&=\sgup\sbup\nabla T-\sgdn\sbdn\nabla T-(\sgup-\sgdn)\nabla\vs
\end{align}
\end{widetext}
Then effective Seebeck coefficient is modified as
\begin{equation*}
S_\mathrm{eff}=\frac{\sgup\sbup+\sgdn\sbdn+\sgpt\sbpt}{\sgco+\sgpt}=\frac{\sgco\sbco+\sgpt\sbpt}{\sgco+\sgpt}
\end{equation*}
Similar with the case of single layer in charge transport, the effective Seebeck coefficient is the summation of weighted Seebeck coefficient of each layer and each spin channel with the weight factor being their conductivity ratio of $\sigma_i/\sum\sigma_i$. However, thermally activated spin transport in the bilayer is remarkably different by the single layer as shown in the second term of right hand side of Eq.\ (S11).
\begin{equation}
\jsco=\frac{(P-\pT)\sgco^2\sbco}{\sgco+\sgpt}\nabla T+\frac{(P\sbpt-\pT\sbco)\sgpt\sgco}{\sgco+\sgpt}\nabla T
\end{equation}
Another contribution to pure spin current arises due to variation of Seebeck coefficient between Co and Pt. To highlight its influence, one could image a fictional system where intrinsic spin current induced by spin polarization of Seebeck coefficient of Co is zero or $\pT=P$. In this system,
\begin{equation}
\jsco=P(\sbpt-\sbco)\frac{\sgpt\sgco}{\sgco+\sgpt}\nabla T
\end{equation}
Diffusive current $\sgco\sbco\nabla T$ driven by temperature gradient completely cancels drift current $\sgco\nabla\vs$ in a single magnetic layer. However, the equilibrium between drift current and diffusion current inside cobalt layer is broken by another adjacent layer of platinum by shielding part of drift current inside the cobalt layer. At this time, the drift current inside the cobalt layer is $\sgco\nabla\vs=\sgco S_\mathrm{eff}\nabla T=\sgco(\sgco\sbco+\sgpt\sbpt)\nabla T/(\sgco+\sgpt)$ while the diffusion current in it still maintains the value of $\sgco\sbco\nabla T$ as in the single cobalt layer.
Therefore a net current $\sgco(\sgco\sbco+\sgpt\sbpt)\nabla T/(\sgco+\sgpt)-\sgco\sbco\nabla T=(\sbpt-\sbco)\sgco\sgpt\nabla T/(\sgco+\sgpt)$ arises in the Co layer, which is accurately balanced by another net current $\sgpt\nabla\vs-\sgpt\sbpt\nabla T=(\sbco-\sbpt)\sgco\sgpt\nabla T/(\sgco+\sgpt)$ in the Pt layer for the whole bilayer system. However the net current in the Co layer is spin polarized while the counterpart in the Pt layer is not. Thus a pure spin current of $P(\sbpt-\sbco)\sgco\sgpt\nabla T/(\sgco+\sgpt)$ is generated inside the bilayer system as Eq.\ (S12) predicts. The first term in Eq.\ (S12) originates from the spin-dependent Seebeck coefficient of Co layer. We name this term as the intrinsic spin-dependent Seebeck effect.The second term relates with variation of normal Seebeck coefficient in different layers, which is named as extrinsic spin-dependent Seebeck effect thereafter.

SOC taken into account, the spin current would also induce an anomalous Nernst voltage. Here we have further supposed that $h$ is smaller enough than a feature length $\lambda$ characterizing length scale where SOC effect of Pt has significant effect on spin current in Co, then
\begin{widetext}
\begin{equation}
\nabla\vn=\frac{\spa\jsco}{\sgco+\sgpt}=\frac{(1-\pT/P)\sigma_\mathrm{Co}^2\sbco+(\sbpt/\sbco-\pT/P)\sgpt\sgco\sbco}{(\sgco+\sgpt)^2}\ha\nabla T
\end{equation}
\end{widetext}

The relation $\ha=P\spa$ has been applied. Then one could obtain Eq.\ (S14) considering relation between Seebeck voltage and temperature gradient.
\begin{widetext}
\begin{equation}
C\equiv\frac{\nabla\vn}{\ha\nabla\vs}=\frac{\frac{\sgco}{\sgco+\sgpt}(1-\pT/P)+\frac{\sgpt}{\sgco+\sgpt}(\sbpt/\sbco-\pT/P)}{1+(\sgpt\sbpt)/(\sgco\sbco)}
\end{equation}
\end{widetext}
It should be pointed out that the above theoretical results (S14) are based on the assumption that $h$ is smaller than $\lambda$. If the condition is not satisfied, the extrinsic spin-dependent Seebeck effect is supposed to disappear.\\ \\
\textbf{C. Estimation of temperature distribution}

Measurement setup is shown in Fig. 1(a) and Fig. 3(a) inset. Heating current is applied along a bar from Pads A to B. Thermal conductance of silicon, [Co/Pt]$_n$ multilayers and silicon dioxide are estimated in the order of $10^{-1}~\mathrm{W/K}$ (lateral thermal transport), $10^{-6}~\mathrm{W/K}$ at most (lateral thermal transport) and $1~\mathrm{W/K}$ (vertical thermal transport), respectively, though their thermal conductivities are nearly in the same order or differ from others only by $1\sim2$ orders of magnitude, $150~\mathrm{W/mK}$ for silicon~\cite{Shanks1963}, $70-100~\mathrm{W/mK}$ for metals~\cite{Terada1999,Wang2014} and $1.5~\mathrm{W/mK}$ for \ce{SiO2}~\cite{Kleiner1996,Goodson1993}. Large difference in thermal conductance of different layers results from large difference in thickness of each layer. Therefore heat flow generated by heating bar tends to first pass through \ce{SiO2}\ layer underneath the heating bar and is then transported through silicon substrate into environment. It is mainly silicon substrate that dominates heat flow transport and that determines temperature distribution along the [Co/Pt]$_n$ superlattices. Another factor affecting the temperature distribution is geometrical dimensions of sample. Core structure for Seebeck and Nernst measurement along $x$ is within $200~\mathrm{\upmu m}$ while dimensions of silicon wafers are $5~\mathrm{mm}\times 5~\mathrm{mm}\times 550~\mathrm{\upmu m}$, and the measurement structure is placed around the middle of the wafer. Considering the large difference in dimensions between the measurement structure and the substrate wafer, we suppose the measurement structure is placed on an infinite large substrate for simplicity in analysis of temperature distribution. Ignoring heat radiation and at steady state, one can obtain from energy conservation law,
  \begin{align*}
  c\rho\uppi r\dif r\dif T &=[k\uppi r T'(r)-k\uppi (r+\dif r)T'(r+\dif r)]\dif t\\
  c\rho r \frac{\dif T}{\dif t} &=-k[T'(r)+rT''(r)]=0
  \end{align*}
Here $c$ is heat capacity of silicon, $T$ is temperature, $k$ is thermal conductivity and $\rho$ is density. The distance away from the heating bar is $r$, as shown in Fig. S1. Thus at steady state,
\begin{equation}
T'(r)=\frac{A}{r}
\end{equation}
where $A$ is a constant and the temperature gradient $\nabla T$ is inversely proportional to $r$, consistent with Ref.\ \cite{Guo2013}. Eq.\ (S15) demonstrates that $\nabla T$ could be extremely large as the point where we measure ANE voltage is close to the heating bar $(r\rightarrow0)$. Being hard to be realized in conventional methods, this large $\nabla T$ could induce a large ANE signal, which helps to improve signal-to-noise ratios in our setup. We set the distance between the point and the edge of the heating bar to be $7~\mathrm{\upmu m}$ .

\begin{figure}[htb!]
{
\includegraphics[width=8.5 cm]{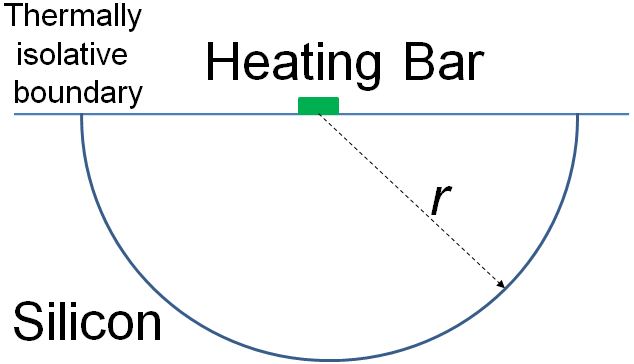}
\caption{\label{figs1}Isothermal line in silicon, temperature at $r$ depends linearly on $1/r$.}
}
\end{figure}

We also conducted finite elementary modeling (FEM) for temperature distribution as shown in Fig.~S2. As predicted by Eq.\ (S15), large $\nabla T$ indeed locates near the heating bar with a very fast decay of $\nabla T$ along $x$. Here all the boundaries were set to be thermally isolative in FEM. The decay of $\nabla T$ along $x$ here is found to be even faster than prediction of Eq.\ (S15). It should be addressed that since $\nabla T$ nonlinearly depends on $x$, $\nabla T$ at Point I $(\nabla T_\mathrm{I})$ where we measured Nernst voltages would be much larger than average temperature gradient between Point I and Point E $((T_\mathrm{I}-T_\mathrm{E})/r_\mathrm{IE})$ between which we measured Seebeck voltages. This is the reason why we have not directly applied Eq.\ (7) or Eq.\ (S14) in our setup to obtain the value of $\pT$. Here, instead, we reasonably assume that similar placement of heating bar on large silicon substrates under the same heating power would result in a similar temperature distribution near the heating bar. Under this assumption, we could thus use [Eq.\ (S14)/Eq.\ (S7)] to cancel the same geometry-dependent factor $(\nabla T_\mathrm{I})/[(T_\mathrm{I}-T_\mathrm{E})/r_\mathrm{IE}]$ . Besides, we also applied $\nn\equiv\dif\vn/\dif P$ and $\ns\equiv\dif\vs/\dif P$ instead of $\dif\vn/\dif\nabla T$ and $\dif\vs/\dif\nabla T$  to indirectly characterize Nernst and Seebeck coefficient since the concrete value of $\nabla T$ is hard to be obtained while $P$ could be accurately controlled.

It is worth reaccentuating that large $\nabla T$, testified by large Nernst voltage about $20~\mathrm{\upmu V}$ in our measurement and hard to be realized in conventional heating setup, could be even further larger in nanodevices where spacing between heating bar and detecting circuit is in the range of several tens of nanometers in nowadays electronics and thus could contribute a higher signal which has easier detectability.
\begin{figure}[htb!]
{\centering
\includegraphics[width=8.5 cm]{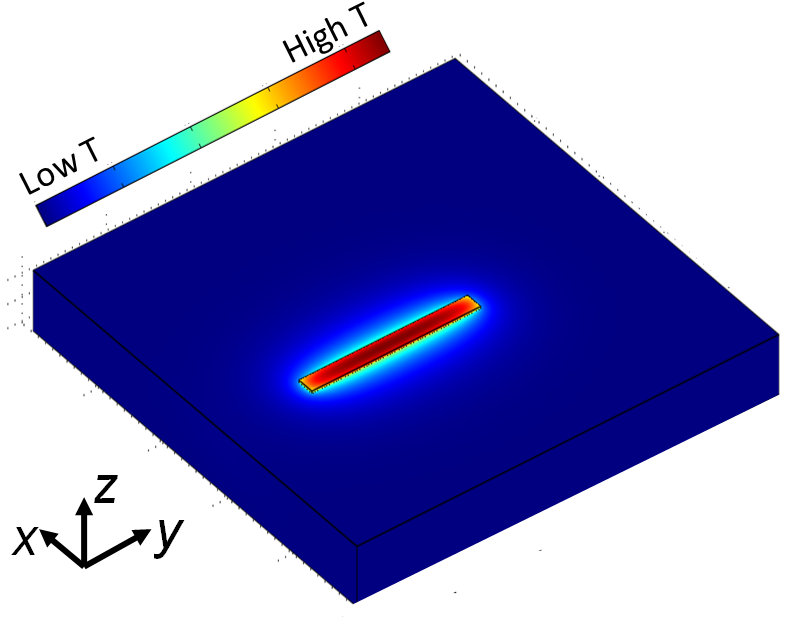}
\caption{\label{figs2}Temperature distribution of silicon on which there exists a heating bar}
}
\end{figure}
\\ \\
\textbf{D. Dependence of Seebeck voltages on heating power}

Seebeck voltages of different samples linearly depended on heating power ($P$) as expected since $\dt$ linearly depends on $P$. Fig.~S3 offset Seebeck voltages of different samples for clarity. Their slopes of $\ns\equiv\dif\vs/\dif P$ are shown as inset in Fig.~S3. Among these samples, Co has largest $\abs{\ns}$ about ($23\pm2$)$\unw$ while Pt has smallest one about ($7.8\pm0.4$)$\unw$. The $\abs{\ns}$ of the superlattices shows no systematic dependence on the number of interface. Their average $\abs{\ns}$ was about ($16\pm3.0$)$\unw$.
\begin{figure}[htb!]
{\centering
\includegraphics[width=8.5 cm]{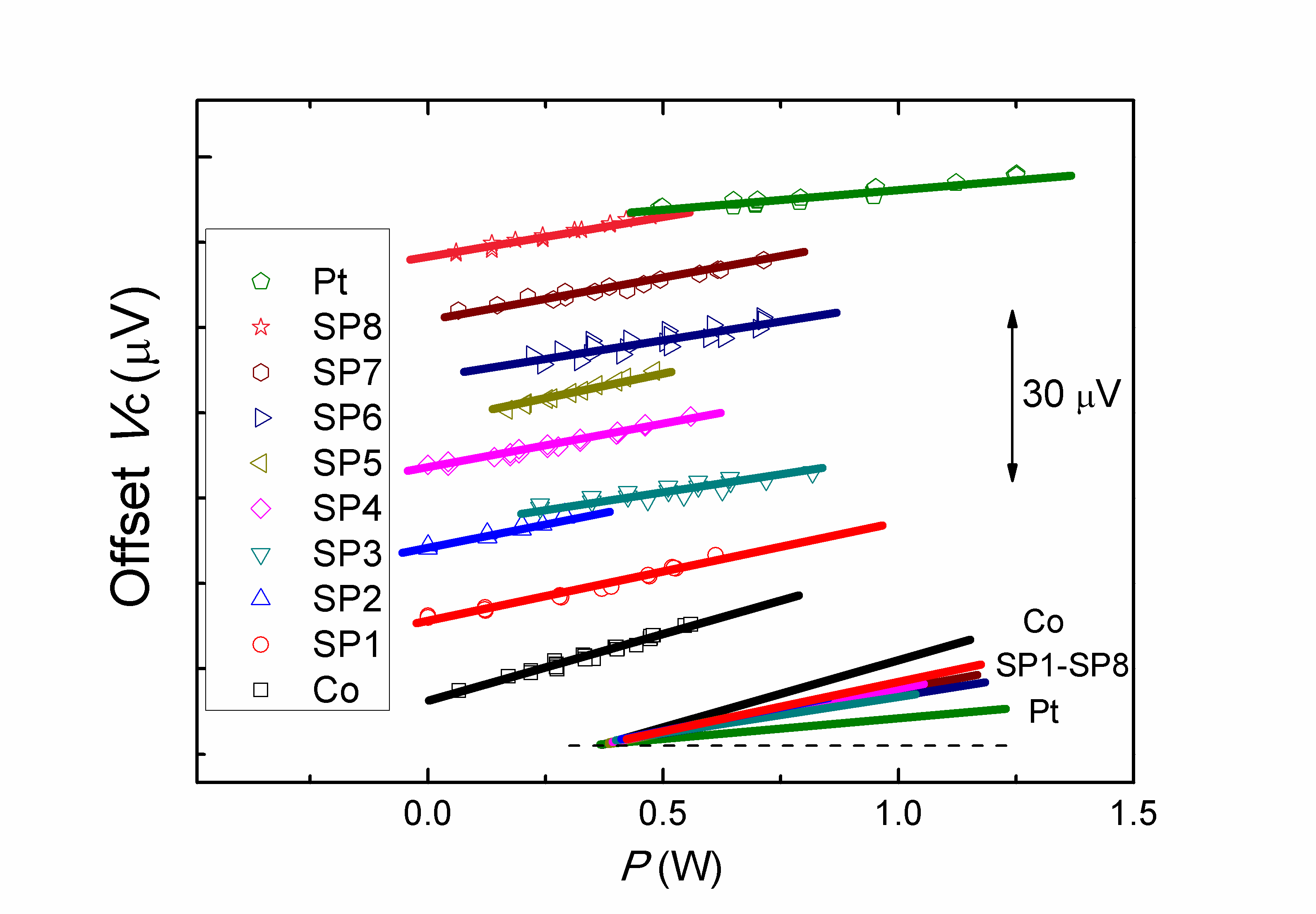}
\caption{\label{figs3}Dependence of offset Seebeck voltage $\vs$ of superlattices on heating power}
}
\end{figure}
\\ \\
\begin{figure}[htb!]
{\centering
\includegraphics[width=8.5 cm]{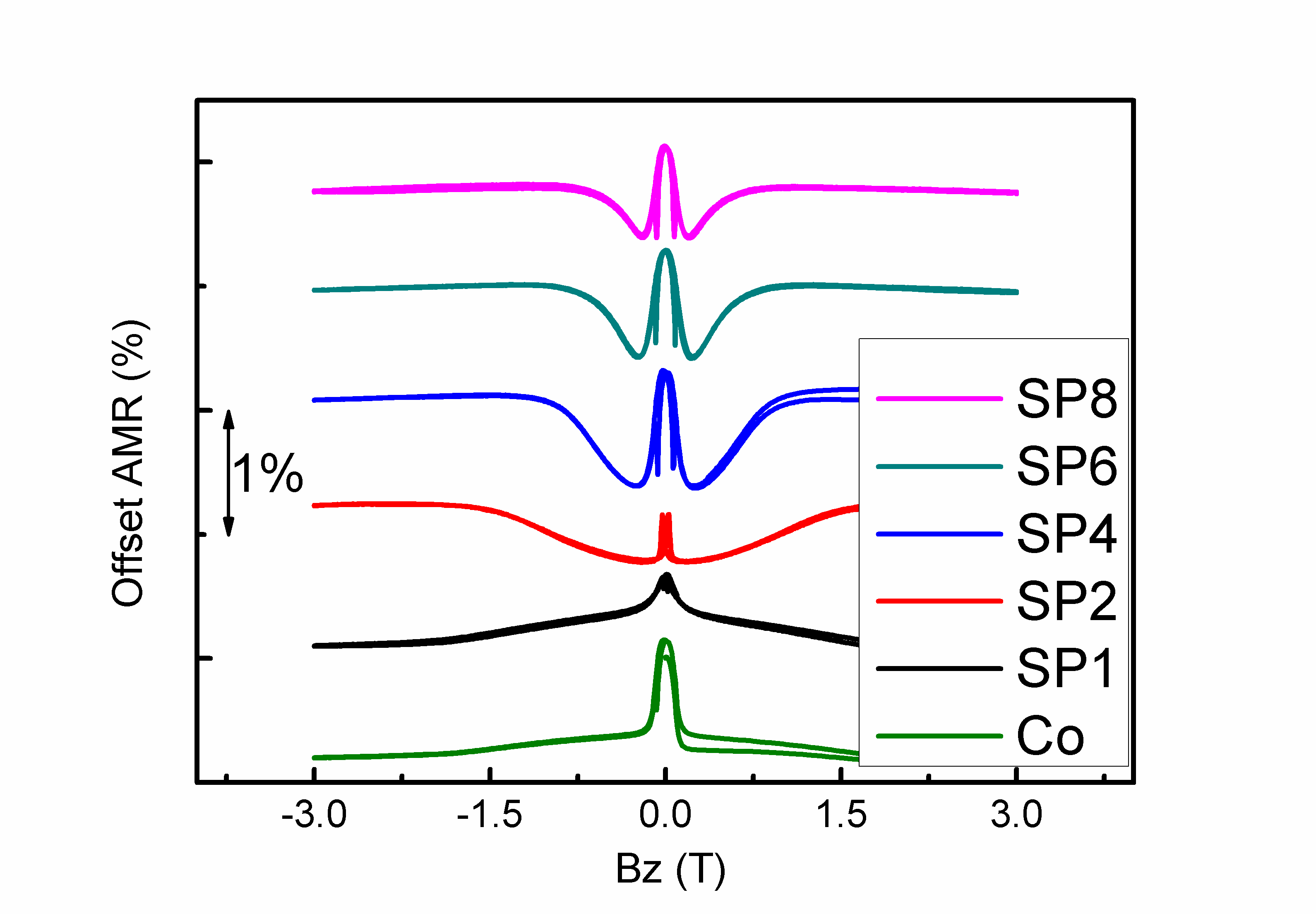}
\caption{\label{figs4}AMR of pure Co and superlattices under magnetic field along $z$ direction}
}
\end{figure}
\\ \\
\textbf{E. Magnetic and transport properties of [Co/Pt]$_n$ multilayers and pure Co layer}

Anisotropic magnetoresistance (AMR) of the superlattices and the Co layer are shown in Fig.~S4. During AMR measurement, magnetic field is applied normal to plane, the same direction as we conduct Nernst measurement and anomalous Hall measurement. The AMR of all the samples are smaller than 1\%, which could introduce about 1\% at most inaccuracy in heating power by $I^2R$. Here $I$ is heating current and $R$ is resistance of the main bar in Fig.~1(a). It should be pointed out that the heating power at $+3~\mathrm{T}$ is the same as that at $-3~\mathrm{T}$ because $R$ is an even function of $B$ at large field. Therefore inaccuracy in final Nernst voltage introduced by the AMR of the heating bar should be even much smaller than 1\% and thus this AMR effect was ignored in our analysis in the main text. The evolution rule of the AMR with increase in period number $n$ is beyond scope of this manuscript and will be discussed elsewhere. The magnetizations of the superlattices are weakly dependent on the period $n$ and their average magnetizations are about $1370~\mathrm{emu/cc}$.

%